\documentclass[10pt]{iopart}
\usepackage{iopams}
\usepackage{graphicx}
\usepackage{placeins}
\usepackage{siunitx}
\usepackage{citesort}

\newcommand{\Yb}{${}^{172}$Yb$^+$}
\begin{document}
	\title[Nanofriction and motion of defects]{Nanofriction and motion of topological defects in self-organized ion Coulomb crystals}
	\author{J. Kiethe$^1$, R. Nigmatullin$^2$, T. Schmirander$^1$, D. Kalincev$^1$, T.E. Mehlst\"{a}ubler$^1$}
	\address{$^1$ Physikalisch-Technische Bundesanstalt, Bundesallee 100, 38116 Braunschweig, Germany}
	\address{$^2$ Complex Systems Research Group, Faculty of Engineering and IT, The University of Sydney, Sydney, New South Wales 2006, Australia}
	\ead{tanja.mehlstaeubler@ptb.de}
	
	\begin{abstract}
		We study nanofriction  in an ion Coulomb crystal under the presence of topological defects. We observe signatures of the pinning to sliding transition, namely the transverse vibrational soft mode signaling the transition and the symmetry breaking of the crystal configuration. We present details on experimental measurements and numerical simulations of this system, including the gaps in the hull function and the spectroscopy of the localized defect mode. We also investigate how external forces, such as those due to anharmonic potentials or differential light pressure, break the intrinsic crystal symmetry, thereby reducing mode softening near the sliding transition. We find that the local structure and position of the topological defect is essential for the presence of the soft mode and illustrate how the defect changes its properties, when it moves through the crystal. 
	\end{abstract}
	\maketitle
	\ioptwocol 

\section{Introduction}\label{sec:Introduction}
Friction influences many natural phenomena over several orders of magnitude of relevant length scales, from earthquakes to biological molecules \cite{Vanossi2013}. Especially nanofriction processes on the atomic scale are nowadays of technical importance, due to advances made in such fields as nanofabrication \cite{Kudernac2011} and biotechnology \cite{Kuehner2007,Ward2015}. To gain insights into the \emph{in-situ} dynamics of nanofriction processes, an emulation via laser-cooled and trapped ions was proposed \cite{Garcia-Mata2007,Benasi2011,Pruttivarasin2011,Mandelli2013,Fogarty2015}. These works suggested emulating one of the most fundamental models of nanofriction, i.e. the Frenkel-Kontorova (FK) model \cite{Braun2013}, using ion chains in an optical lattice \cite{Aubry1983,Sharma1984,Braiman1990}. Following the proposals, Bylinskii \textit{et al.} were able to trap 5 ions in an optical lattice, demonstrating the onset of reduced friction \cite{Bylinskii2015},  the velocity dependence of the stick-slip motion \cite{Gangloff2015}, as well as single ion multi-slip behavior \cite{Counts2017}. Furthermore, they observed structural symmetry-breaking at the pinning to sliding transition \cite{Bylinkskii2016}, an Aubry-type (AT) transition.

In actual nanocontacts, two layers of atoms are interacting with each other. The corrugation is then no longer given by a rigid potential, but rather by a deformable layer \cite{Matsukawa1994}. Recently, we presented a self-organized system, consisting of two deformable back-acting ion chains, without a fixed corrugation potential \cite{Kiethe2017}. We showed, that two-dimensional ion Coulomb systems under the presence of a topological defect \cite{Mielenz2013,Pyka2013,Ulm2013,Ejtemaee2013} exhibit features of an Aubry-type transition, namely the breaking of analyticity of a hull function and a soft phonon mode. We also observed the symmetry breaking inside the crystal, which is a typical signature of finite systems \cite{Sharma1984,Braiman1990}. Our system shows similarities to other self-organized sliding systems, such as fibrous composite materials \cite{Ward2015}, sliding DNA strands \cite{Kumar2010} and propagation of protein loops \cite{Sieradzan2014}. While the investigation of nanofriction with the help of discrete topological defects, also known as kink solitons, is a new approach, these defects have been extensively studied with respect to the Kibble-Zurek mechanism \cite{Mielenz2013,Pyka2013,Ulm2013,Ejtemaee2013,DelCampo2010,Silvi2016,DeChiara2010} and to quantum information \cite{Landa2010,Landa2014}. Also, the directed transport of topological defects via a ratchet mechanism has been recently reported \cite{Brox2017}.

In the following we discuss in more detail our experimental and numerical findings of this tribological system. We explain our approach to the calculation of the hull function in self-organized systems and discuss secondary gaps in the hull functions due to the inhomogeneity of the charge density. We detail our spectroscopic method, which was used to probe the localized kink soliton mode, transition from sticking to sliding. We also study the influence of temperature on the vibrational spectra using molecular dynamics simulations and Fourier transformation techniques. We numerically investigate external forces that break the local crystal symmetry, and influence the mode frequencies of the system. Forces under investigation result from effects, such as anharmonicities of the axial trapping potential, axial micromotion and laser light pressure. We show, that the position of the topological defect is sensitive to forces on the \SI{e-21}{\newton} level. We numerically investigate the motion of extended topological defects through the crystal induced by differential light forces, and discuss the associated periodic change in their physical properties, such as local structure and mode frequency.

This paper is structured in the following way. \Sref{sec:ATT_in_self_organized_systems} introduces our experimental system and describes the observation of two main features of the Aubry-type transition in finite atomic systems, namely the symmetry breaking and the soft vibrational mode. Then we introduce the definitions of an order parameter and the hull functions, and how they can be applied to self-organized systems. In \Sref{sec:Environment} we analyze external forces that could break the intrinsic crystal symmetry. Building on these results, in \Sref{sec:KinkMovement}, we study the behavior of the topological defect, as it is moved through the crystal. Finally, in \Sref{sec:Discussion}, we discuss our results and give an outlook to future experiments and theoretical investigations.

\section{Pinning to sliding transition in a self-organized system}\label{sec:ATT_in_self_organized_systems}
\begin{figure}
	\centering
	\includegraphics[width=0.45\textwidth]{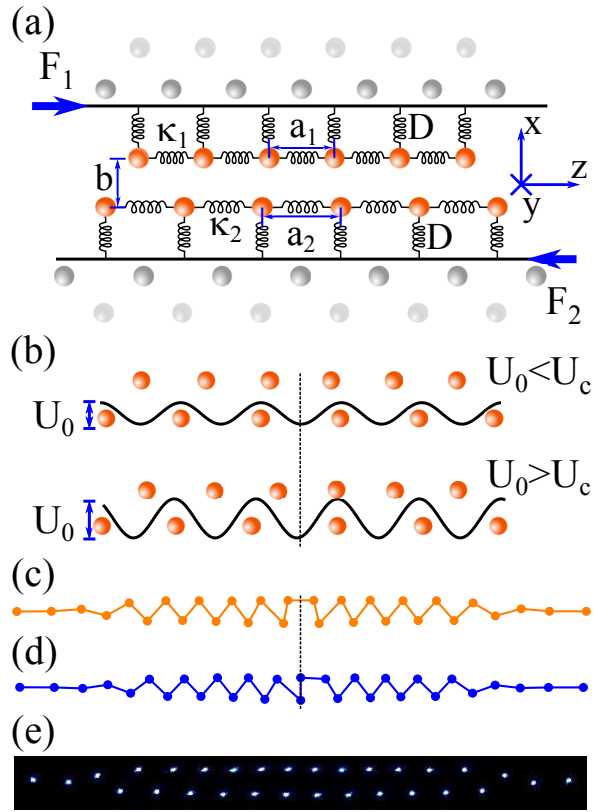}
	\caption{Nanofriction model system in a self-organized ion Coulomb crystal. (a) The central part of a Coulomb crystal can be seen as a back-acting interface between two atomically flat surfaces. Connection to a rigid solid (in our system the trap) is modeled with springs of strength $D$. Particles in one chain interact via springs of strength $\kappa_{1/2}$, which depend on the intra chain spacing $a_{1/2}$. The distances are not identical, due to the topological defect. The inter chain distance $b$ determines the corrugation. Sliding of the model can be emulated by pushing one chain with force $F_1$, and pushing the other with force $F_2$. The coordinate system defines the axial ($z$) and radial ($x$) direction. (b) Symmetry breaking above the critical corrugation strength $U_c$. Decreasing the distance $b$ between the chains, increases the corrugation until it crosses $U_{c}$ and the defect moves away from the center (dashed line). (c) Horizontal kink. The two inner most ions are in the same chain. The number of ions in the upper and lower chain is different. The dashed chain shows the symmetry axis. (d) Vertical Kink. The two center ions are in different chains. The number of ions per chain is identical. (e) Experimental image of a horizontal defect. Exposure time is \SI{700}{\milli\second}. The $xz$ plane of the crystal has an angle of approximately \SI{40}{\degree} to the CCD chip, resulting in a reduced observed radial distance between the chains. Graphics (a) and (b) are from \cite{Kiethe2017}.}\label{fig:model_system}
\end{figure}

Our system consists of a two-dimensional ion Coulomb crystal in the zigzag phase with an extended topological defect in the center. The frictional process of interest is the sliding of the two chains against each other. In the following, we focus on the axial vector components of the motion, reducing the system to a one-dimensional problem. The interaction energies within the chains, $U_{\text{intra}}$, and between the chains, $U_{\text{inter}}$, characterize the regime of friction. To determine the critical point, we can estimate these energies from the characteristic distances in the Coulomb crystals. To do so, we model the two linear chains as classical particles of mass $m$, which interact within each chain via springs with constant $\kappa_{i}$, where $i$ is $1$ or $2$, identifying the chains as illustrated in \Fref{fig:model_system}(a). The spring constants of the chains are different, due to the topological defect introducing a small mismatch between the inter-ion distances $a_i$. The chains are separated by a distance $b$ and all masses are connected to a rigid support by springs of stiffness $D=m\omega_{z}^2$, which corresponds to the axial confinement of the ion trap $\omega_{z}$. Moving the chains against each other can be done via light forces acting on each chain independently, which we name $F_i$. Given this simplified model we can employ a harmonic approximation to roughly estimate the interaction energy of ions within the top chain as $U_{\text{intra}}\approx\frac{1}{2}\kappa_{1}z^2$. The energy scale of the corrugation by the other chain can locally be approximated as $U_{\text{inter}}\approx\frac{1}{2}U_0\left[\cos\left(2\pi a_2^{-1}z\right)+1\right]$, where $U_0$ is the strength of the corrugation, which depends on $b$. If the chains are brought closer together $U_0$ increases, changing the relative strength of the interactions. At a certain distance a critical depth of the corrugation potential $U_{c}$ is crossed and the system changes from the sliding to pinning regime. At this point we expect a symmetry breaking of the crystal structure as illustrated in \Fref{fig:model_system}(b). The transition should happen when the competing energies are in balance. Representing these energies as angular frequencies in the harmonic approximation: ${\omega_{\text{pinning}}=\sqrt{2\pi^2U_0\left(ma_2^2\right)^{-1}}}$ and ${\omega_{\text{natural}}=\sqrt{\kappa_{1}m^{-1}}}$ we can define a dimensionless corrugation parameter 
\begin{equation}
	\eta=\frac{\omega_{\text{pinning}}^2}{\omega_{\text{natural}}^2}.
\end{equation}\label{eq:corrugation}
At the transition we therefore expect an $\eta$ of order $1$.

In order to estimate $\eta$, we approximate the frequencies $\omega_{\text{pinning}}$ and $\omega_{\text{natural}}$, which describe a many-body interaction, with the help of the potentials, seen by one of the central ions. In a two-dimensional Coulomb crystal with a topological defect shown in \Fref{fig:model_system}(b), both central ions are located on the slope of $U_{\text{inter}}$. As a harmonic approximation is not valid there, we use the interaction energies $\tilde{U}_{\text{inter}}$ and $\tilde{U}_{\text{intra}}$ of a zigzag without defect as a further approximation. In such a crystal, all ions of one chain are located in the potential minimum of $\tilde{U}_{\text{inter}}$. The second-order Taylor expansion of $\tilde{U}_{\text{inter}}$ and $\tilde{U}_{\text{intra}}$ are than proportional to the approximate frequencies $\tilde{\omega}_{\text{pinning}}$ and $\tilde{\omega}_{\text{natural}}$, respectively. This will result in an approximate corrugation parameter $\tilde{\eta}$, which has a critical value $\tilde{\eta}_{c}$ different from 1.

\subsection{Experimental system}\label{ssec:experiment}	
In this section we detail the experimental parameters necessary for implementing the above model system. We trap an ion Coulomb crystal with 30 \Yb ions in a high-precision linear Paul trap \cite{Keller2018}. The ion trap chips are laser-machined with low tolerances and carefully aligned to reduce symmetry breaking of the electrode configuration. This reduces unwanted odd numbered anharmonic terms in the axial potential seen by the ions. The ions are laser-cooled on the strong ${}^2$S$_{1/2}\rightarrow{}^2$P$_{1/2}$ dipole transition  to temperatures in the low \si{\milli\kelvin} regime. The laser has a central wavelength of \SI{369.5}{\nano\metre} and a beam diameter of \SI{2.56}{\milli\metre} in axial ($z$) and \SI{80}{\micro\meter} in radial ($\approx\sqrt{\left(x^2+y^2\right)}$) direction. Typical laser powers are \SIrange{200}{300}{\micro\watt}. The fluorescence of the ions is imaged onto an electron multiplying (EM) $512\times 512$ pixel CCD camera via a lens system with $\text{NA}=0.2$ and magnification $24$. The crystals $xz$ plane has an angle of roughly \SI{40}{\degree} to the CCD chip. This reduces the observed radial distance between the ions. On the camera we can resolve individual ions and for an exposure time of \SI{700}{\milli\second} an ion position in a pure zigzag configuration can be resolved to \SI{40}{\nano\metre} by fitting a Gaussian profile to the image. If the system is close to the Aubry-type transition, the resolution for the central ions, which are part of the topological defect, worsens.

In order to confine the crystal to the two-dimensional $xz$ plane, we lift the degeneracy between the two radial potentials with the help of the dc potentials \cite{Herschbach2012,Keller2018}, achieving a radial anisotropy of $\omega_x/\omega_y=1.3$. In a two-dimensional crystal, the observed structural phase depends only on the relative magnitude of the transverse and axial confinement, quantified by the ratio of the respective secular frequencies $\alpha\equiv \omega_x/\omega_y$. This parameter is used in our calculations and experiments as a control to change the distance $b$ and subsequently the corrugation $\eta$. For $4.81<\alpha<13.6$ a crystal with 30 ions will be in the two-dimensional zigzag phase \cite{Landa2013}.

For a fixed axial secular frequency $\omega_z\approx 2\pi\cdot\SI{25}{\kilo\hertz}$, topological defects can be created by rapidly quenching the rf amplitude of the radial confinement \cite{Pyka2013}. In terms of radial secular frequencies, this quench changes the confinement from ${\omega_x\approx 2\pi\times\SI{500}{\kilo\hertz}}$ to values ranging from ${2\pi\times\SI{140}{\kilo\hertz}}$ to ${2\pi\times\SI{200}{\kilo\hertz}}$ in \SI{58}{\micro\second}. During this quench, several types of topological defects can be created in a finite ion Coulomb crystal in the two-dimensional phase \cite{Landa2013}. The defect type is determined by the control parameter $\alpha$. For an ion number ${N=30}$ and the trap ratio in the range of $4.81<\alpha<7.76$, extended kinks are observed. For $\alpha>7.76$, the defects are of the odd type \cite{Partner2013,Landa2013} and for $\alpha<4.81$ the crystal exhibits a three-dimensional helix configuration \cite{Nigmatullin2016}. In the extended defect regime, which is relevant to the current paper, two kinds of defects can be observed. We refer to the these defects as horizontal (H) kink and vertical (V) kink. The two kinks are depicted in \Fref{fig:model_system}(c) and (d). The most prominent difference between them is the position of the two inner most ions. In the horizontal case these ions are in the same chain and occupy the same minimum of the Coulomb potential produced by the opposite chain. In the vertical case these ions are in different chains and the ion of the upper chain lies on the potential maximum created by of the lower chain. Another difference is the number of ions in each chain. Ignoring the first 4 ions on the left and right side of a crystal, as they are still close to the linear chain, a crystal with a V kink displays equal ion numbers for both chains. A crystal with an H kink exhibits a two ion difference between the chains. The kink is found in the center of the crystal, due its inhomogeneity \cite{Partner2013}. In our experiments, both defects are created in a rf quench and are initially of odd type. They change to the extended type when the rf amplitude is lowered further \cite{Partner2013,Partner2015}. The overall probability to create a defect is around \SI{30}{\percent}, but only the H defect is used in the experiments, because only this defect shows a soft mode, if it is unperturbed, as we will show in \Sref{ssec:Spectroscopy}. Therefore, roughly \SI{15}{\percent} of all quenches result in a crystal with the necessary extended defect. An experimental image is shown in \Fref{fig:model_system}(e). The observed crystal configurations are highly symmetric due to the well controlled environment provided by the ion trap.

\subsection{Structural symmetry breaking}
For a finite system a symmetry breaking is expected at the Aubry-type transition \cite{Sharma1984,Braiman1990}.
\begin{figure*}
	\centering
	\includegraphics[width=0.9\textwidth]{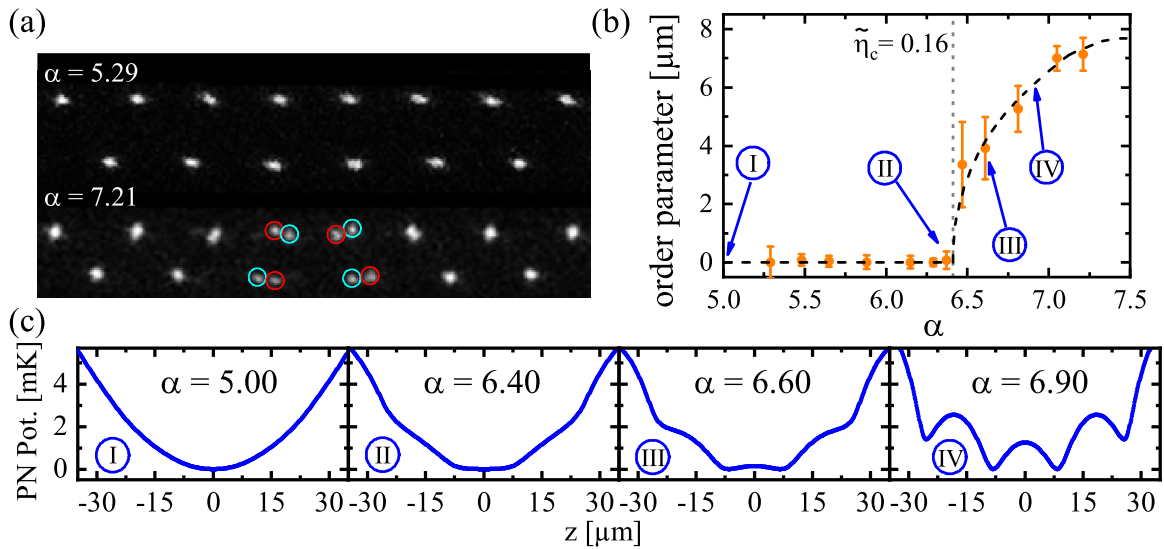}
	\caption{Symmetry breaking at the pinning to sliding transition. (a) Experimentally observed crystal configurations below the Aubry-type transition, at $\alpha=5.29$, and above, at $\alpha=7.21$. The images are taken with \SI{700}{\milli\second} exposure time. The crystal is at a finite temperature in the \si{\milli\kelvin} regime, which leads to multiple fluorescence spots above $\alpha_c=6.41$, as the topological defect moves between the different stable positions. An angle of \SI{40}{\degree} between the crystals $xz$ plane and the CCD chip, results in apparent smaller radial distances. (b) Absolute order parameter $\left|\Phi_{\text{red}}\right|$ against the trapping ratio $\alpha$. Experimental data (orange circles) and numerically obtained values for ${T=\SI{0}{\kelvin}}$ (black dashed line) are shown. Experimental data represents a weighted average over $5$ to $26$ measurements per point, with exception of $\alpha=7.21$, where only 2 configurations were observed. Error bars are given as one standard deviation weighted by fit errors. The critical point is $\alpha_c\approx6.41$ (approximate critical corrugation parameter $\tilde{\eta}_c\approx0.16$). (c) Peierls-Nabarro potential for different trapping ratios $\alpha$. The increasing PN barriers lead to the symmetry breaking of the system. Barriers for $\alpha>\alpha_c$ are on the order of the temperature in the system, leading to multiple observed ion positions as the defect can move between different minima.}\label{fig:SymBreak}
\end{figure*}
 In the experiment we observed several fluorescence spots in the symmetry broken regime, as can be seen in \Fref{fig:SymBreak}(a) for ${\alpha=7.21}$, where fluorescence spots belonging to the same configuration are indicated by circles. The symmetry breaking is a direct result of the emerging barriers in the Peierls-Nabarro (PN) potential \cite{Braun2013} at the critical value of the control parameter, $\alpha_c$. This potential describes the collective dynamics of a discrete non-linear system. In our case, the PN barrier rises in the center of the crystal, see \Fref{fig:SymBreak}(c). As a result, the extended defect destabilizes and slips into the adjacent minimum of the PN potential, which alters the ion positions and breaks the axial mirror symmetry of the crystal. The finite temperature of the ions results in switching between stable configurations, and hence multiple spots per ion are observed during a finite exposure time of a few tens of \si{\milli\second}. The barrier between the minima in the PN potential are on the order of \SIrange{1}{2}{\milli\kelvin}, comparable to the temperature of the crystal, which we estimate to be in the low \si{\milli\kelvin} regime \cite{Partner2015}. We quantify the symmetry breaking by choosing an order parameter $\Phi$, defined as the sum of the relative axial differences between next-neighbor ions from different chains
\begin{equation}\label{eq:Phi}
\Phi=\sum_{i\in \text{Chain }1}\text{sgn}\left(z_i\right)\cdot\min_{j\in \text{Chain }2}\left|z_i-z_j\right| ,
\end{equation}
where $z_i$ is the axial coordinate of the $i$th ion and $z=0$ is the axis of symmetry for $\alpha<\alpha_c$. Below the transition the order parameter is $0$. From numerical simulations, which are described in detail in \Sref{sssec:NumericalCalculations}, we extract stable crystal configurations, which yield the order parameter shown in the dashed line of \Fref{fig:SymBreak}(b). The critical point is $\alpha_c\approx 6.41$, at which $\Phi$ exhibits a sudden cusp. The corresponding approximate critical corrugation parameter $\tilde{\eta}_c$ is $0.16$, which is different from $1$, since it is only a rough estimate.
Taking fluorescence images of crystals at various $\omega_x$, we extract the ion positions via fitting the fluorescence profile of the individual ions to a Gaussian function. To avoid errors due to optical aberrations far away from the optical axis and to reduce accumulated fitting errors, we do not use the full version of \Eref{eq:Phi}. The reduced order parameter $\Phi_{\text{red}}={\Phi_{i=N/2}+\Phi_{i=N/2+1}}$ only includes the two central terms, which have the largest contribution to $\Phi$. The value of $\alpha_c$ is not influenced by the choice of $\Phi$. The experimental results are plotted as orange circles in \Fref{fig:SymBreak}(b), and they agree well with the numerical calculations. The error bars are one standard deviation weighted by the fit errors. Measuring $\Phi$ for $\alpha>\alpha_c$ close to the critical point is limited by the resolution of the EMCCD camera, as the fluorescence spots of two configurations need to be at least $2\,$pixel apart in order to be identified as distinct configurations.

\subsection{Derivation of the hull function in a self-organized system}
The symmetry breaking in finite systems is also accompanied by the more general breaking of analyticity of the hull functions \cite{Sharma1984}. In the following paragraphs we will discuss the hull function and its numerical calculation in our model system in more detail.

In the Frenkel-Kontorova model the hull function parameterizes the reachable ground state configurations of an incommensurate system \cite{Aubry1983,Braun2013}. If the system is below the Aubry or Aubry-type transition, i.e. free sliding, this function is continuous and analytic. Intuitively this means, that during a sliding process of the chain over the potential, all individual particle positions are a realization of a ground state of the system. This changes when the Aubry transition is crossed, and the system start to exhibit stick-slip motion. Certain positions are unstable and not part of the set of ground state solutions, resulting in a discontinuous hull function with broken analyticity \cite{Aubry1983}. This property of incommensurate sliding systems exists in both infinite and finite systems, though the two cases are not identical.
\begin{figure*}[tpb]
	\centering
	\begin{tabular}[t]{cc}
		\includegraphics[width=0.4\textwidth]{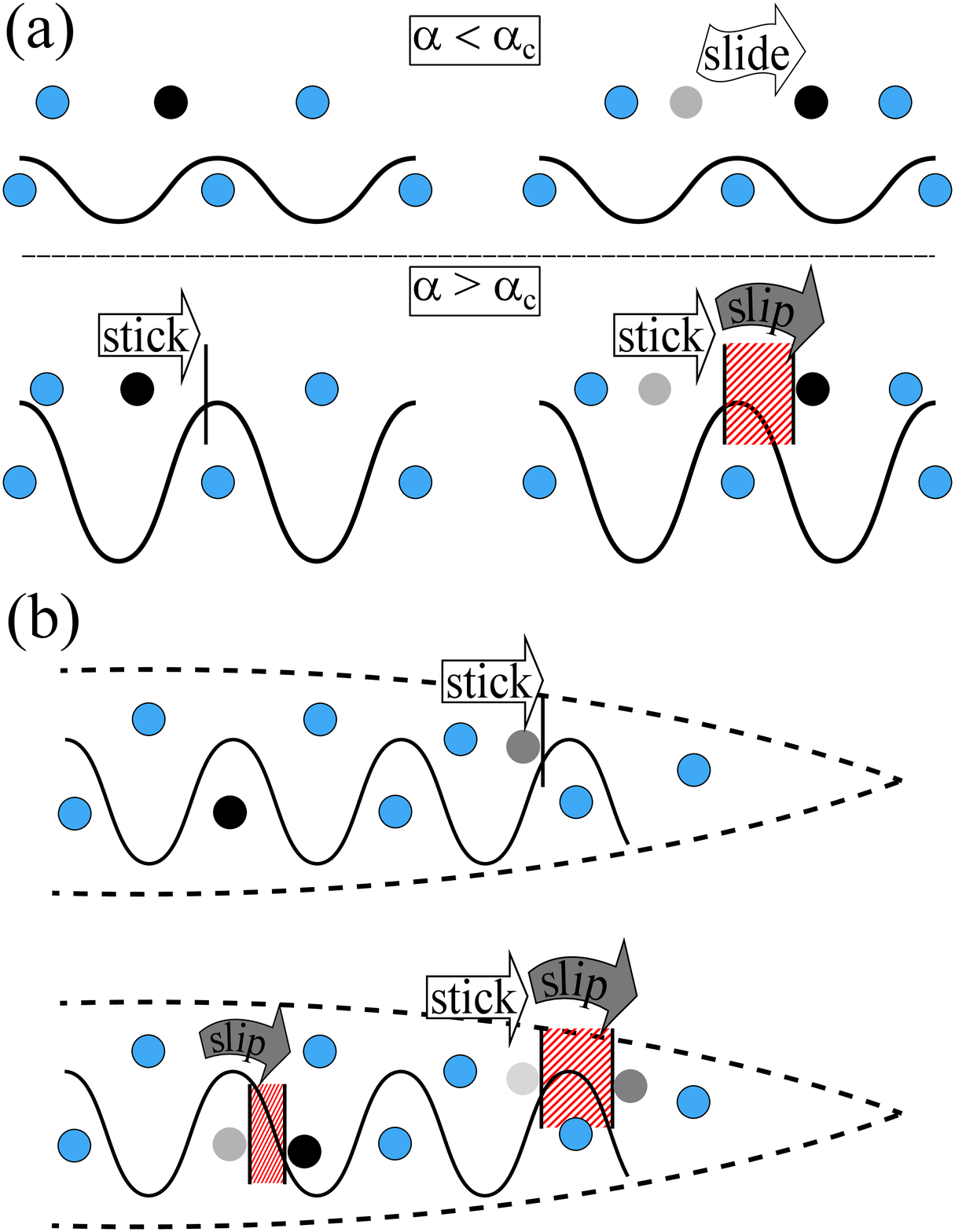}&
		\includegraphics[width=0.4\textwidth]{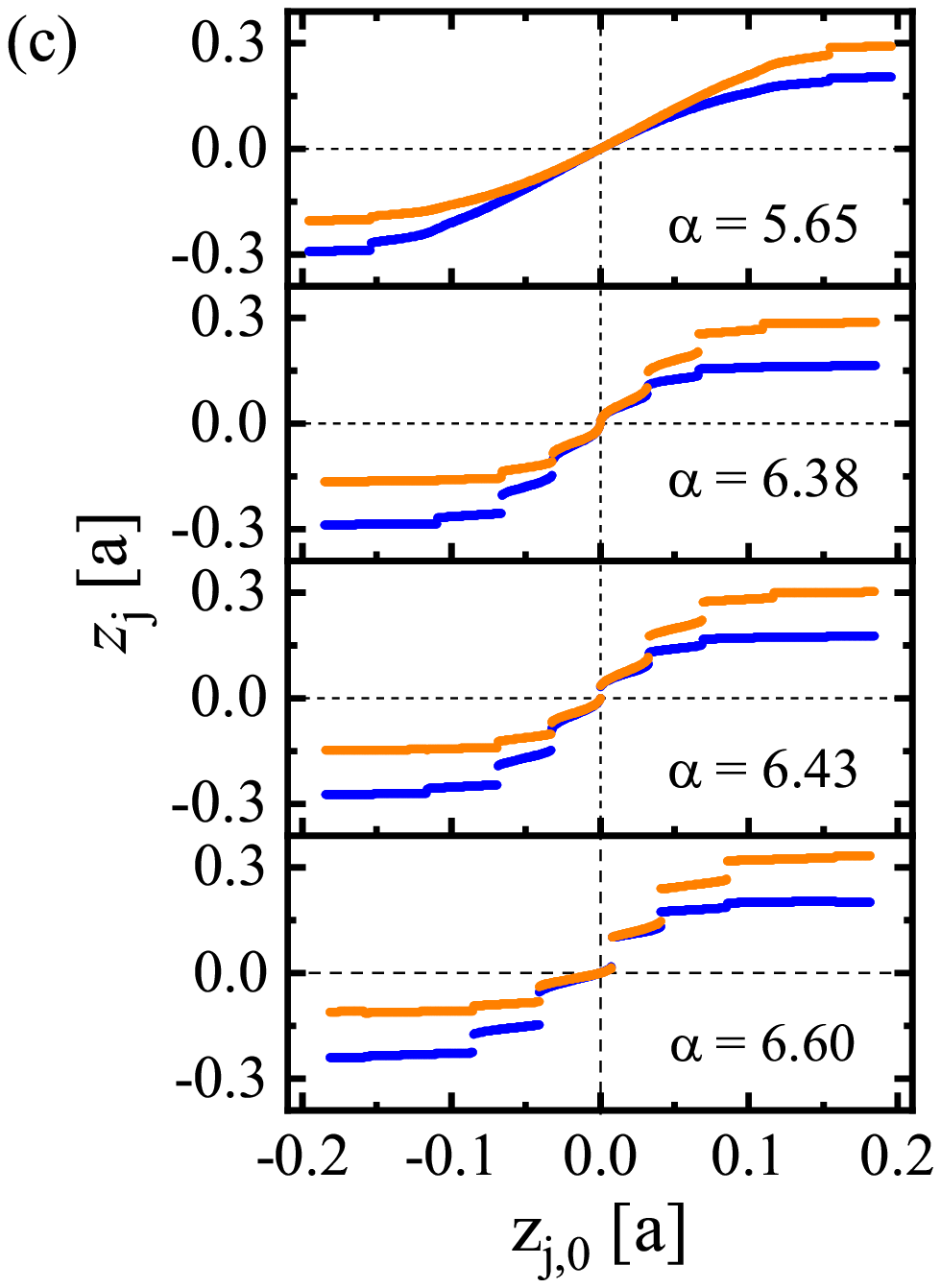}
	\end{tabular}
	\caption{Hull function of an inhomogeneous crystal. (a) Schematic explanation of a primary gap in the hull function. For $\alpha<\alpha_c$ a particle (black circle) can assume all positions, when being pushed. In the hull function no gap is observed. For $\alpha>\alpha_c$ the corrugation is stronger, and several positions are unstable. A slip occurs and a gap appears in the hull function. (b) Secondary gap in the hull function. Due to the Coulomb interaction, a slip of an outer ion (dark gray circle) changes the observed inner ion (black) position suddenly and a gap in the hull function appears. In an inhomogeneous system this can also happen, if $\alpha<\alpha_c$, because the chains are closer together near the crystal edge. (c) Numerically calculated hull functions. The functions are shown for the $15$th (orange) and $16$th (blue) ion for different values of the control parameter $\alpha$. The gaps for $\alpha=5.65$ are due to the topological defect leaving the crystal. The inner gaps for $\alpha=6.38$ are due to stick-slip events away from the center. The primary gap opens up in the center of the hull functions for $\alpha>\alpha_c=6.41$. All positions are in units of the pseudo lattice constant $a$, the distance between the $14$th and $17$th ion.}\label{fig:hullfunctions}
\end{figure*}

In the classical FK model the corrugation potential is not influenced by the particle chain residing above it, which leads to a straightforward representation of the hull function. It is defined as an implicit function $z_j(z_{j,0})$, where $z_j$ is the position of $j$th ion in the presence of the corrugation potential and $z_{j,0}$ is the position of the $j$th ion in the absence of the corrugation potential \cite{Braun2013}. The particle positions can be calculated with the help of a force $F$, that pushes the chain along the axial direction. Doing this calculation with and without a corrugation potential yields $z_j$ and $z_{j,0}$, which determine the implicit relation $z_j\left(z_{j,0}\right)$. Such a procedure is possible, because the corrugation strength $U_0$ can be set to zero without altering the interaction between chain particles. For a self-organized system, like a two-dimensional ion Coulomb crystal, this approach does not work, because the corrugation potential is part of a back-acting system. Removing it without changing the dynamics of the interaction between particles of the same chain is impossible.

However, in a harmonically trapped system, one can push the complete crystal with a force $F$, leaving the local ion interactions unchanged, to obtain a relation between the force and the unperturbed coordinates, ${z_{j,0}\left(F\right)}$. Applying an axial shear force $\Delta F$ onto the chains, slides the particles over the corrugation potential, leading to perturbed positions $z_{j}\left(F\right)$. The shear force is such that one chain is pushed by $+F/2$, while the other chain is pushed by $-F/2$. Then the relative force between the chains is equal to $F$, which was used to push the whole crystal. The unperturbed and perturbed coordinates are related via the applied force $F$, providing a representation of the hull function $z_j\left(z_{j,0}\right)$. The hull functions for the two ions at the center of the crystal, ion 15 and 16, are shown in \Fref{fig:hullfunctions}(c).

In the hull function various gaps open up above the transition due to stick-slip events. These events result from the emergence of the PN barriers above the AT transition, which destabilize crystal configurations, as can be seen in \Fref{fig:SymBreak}(c). The central gap seen in \Fref{fig:hullfunctions}(c) for $\alpha>6.41$ is a direct result of a slip event of the $15$th or $16$th ion, see \Fref{fig:hullfunctions}(a). If the slip event takes place at the position of the defect, the opening in the hull functions is called the primary gap. The non-central gaps in the hull functions of the $15$th and $16$th ions are either remnants of a slip event of other ions or the defect leaving the crystal. If the defect leaves the crystal, a part of the crystal will reorient itself affecting the positions of the central ions. This is observed, for instance, far below the transition at $\alpha = 5.65$ and $z_{j,0}\approx 0.15a$, where $a$ is the axial spacing between the $14$th and $17$th ion, which we consider as a pseudo lattice constant for the crystal center. As the pure zigzag has mostly matched chains, no further information about the AT transition can be gained for bigger forces. Here we note, that, even though the individual ions moved less than a third of a lattice period ($\Delta z< \SI{3}{\micro\metre}$), they have rearranged in a way such, that the defect already left the crystal, which means it moved more than \SI{50}{\micro\metre}. This illustrates why charge transport is facilitated by topological defects \cite{Vanossi2003}.

For $\alpha = 6.38$, still below $\alpha_c$, more than one gap can be seen in the hull function. One in each function is due to the defect leaving the crystal. The others are secondary gaps due to stick-slip events, even though the system is globally in the sliding regime. The stick-slip events occur, because of the inhomogeneity of a harmonically trapped crystal, which reduces the charge density towards the edge of the crystal, as illustrated in \Fref{fig:hullfunctions}(b). This increases the corrugation depth between the ion chains, as the axial distance between ions grows, while the radial distance between the chains decreases. Therefore, the critical point is locally already crossed, if the local ordering is disturbed by a topological defect. The slip events from these ions are seen by the central ions $15$ and $16$, which gain gaps in their hull functions as a result. In a crystal with equidistant ions, the secondary gaps will only exist above $\alpha_c$, as only this regime stick-slip motion is present.

\FloatBarrier
\subsection{Motional mode frequency of the topological defect}\label{ssec:Spectroscopy}
The localized vibrational mode of the topological defect drives the AT transition \cite{Kiethe2017}. The lower its frequency, the lower the energy required, to move the chains against each other. This defect mode is one of the gapped modes, previously considered for quantum information \cite{Landa2010}. Excitation of the high energy gapped was shown in \cite{Brox2017}.  Here we are interested in the low energy gapped mode and how the mode frequency depends on the corrugation strength of the sliding system.

\subsubsection{Spectroscopy}

\begin{figure}
	\centering
	\includegraphics[width=0.45\textwidth]{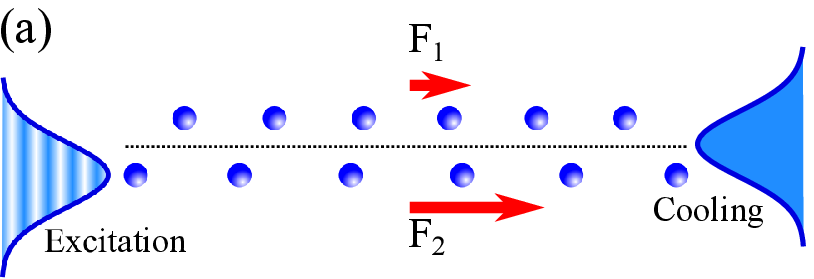}
	\par
	\vspace{0.25cm}
	\includegraphics[width=0.45\textwidth]{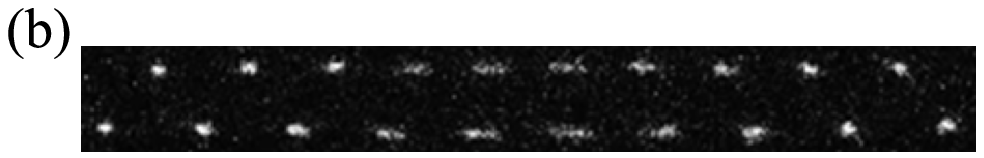}
	\caption{Vibrational mode spectroscopy. (a) Excitation method. The ion crystal is illuminated by a centered laser beam, which continuously cools the ions. A second, axially focused laser beam is centered on one chain, in order to realize a difference in the light forces acting on the chains. The laser is amplitude modulated. (b) Excited shear mode. CCD image with an excited localized defect mode at $\alpha=5.77$. Exposure time is \SI{100}{\milli\second} and cooling laser power of \SI{300}{\micro\watt}. The power of the focused laser was modulated with a frequency of \SI{12.1(3)}{\kilo\hertz} between \SI{0}{\micro\watt} and \SI{35}{\micro\watt}.
	}\label{fig:expSpectroscopy}
\end{figure}

Experimentally we measure the vibrational mode frequency via amplitude modulation of a laser addressing the ${}^2$S$_{1/2}\rightarrow{}^2$P$_{1/2}$ transition. This exerts a periodic force
\begin{equation}
F_p=\frac{F_0}{2}\left[\cos\left(2\pi\cdot\nu t\right)+1\right]
\end{equation}
 on the crystal, where $F_0/2$ is the amplitude of the modulation, which depends on the power of the laser and the detuning from the atomic resonance, and $\nu$ is the modulation frequency. $F_p$ can excite a motional mode $k$, if $\nu$ is close to a mode frequency $\nu_k=\omega_k/\left(2\pi\right)$. Additionally, the intensity distribution of the laser $\bi{A}$ must overlap with the mode vector ${\bmu}_k$. This means, that the scalar product $\bmu_k\cdot \bi{A}$ has to be non-zero. The laser used in our experiments has an angle of \SI{25}{\degree} to the crystal axis and an approximate angle of \SI{50}{\degree} to the $xz$ plane of the crystal. The beam has a Gaussian beam profile with waist of roughly \SI{80}{\micro\metre} in both directions. If the beam is centered on the crystal, the intensity distribution will be identical for both chains. This would result in a zero overlap with the localized defect mode of the crystal, as both chains are excited with the same force and in the same direction. If the intensity maximum is focused radially on one of the chains, see \Fref{fig:expSpectroscopy}(a), then an intensity difference between the chains exist, which results in a non-zero scalar product, $\bmu_k\cdot \bi{A}$, enabling the excitation of this mode. To determine the resonance, $\nu$ is scanned at a rate of \SIrange{1}{2}{\kilo\hertz\per\second} and the spread of the fluorescence of the central ions is recorded with the EMCCD camera. An example image of the excited defect mode is shown in \Fref{fig:expSpectroscopy}(b). Only the central ions move with a significant amplitude, illustrating that the excitation is localized on the defect. The resonance is measured with an uncertainty of \SIrange{300}{400}{\hertz}, which can be improved by a slower frequency scan rate and recording more images per scan. The applied amplitude $F_0/2$ on the crystal needs to be big enough to actually excite observable motion but also has to be small enough, to limit non-linear excitations. In our experiment we found that an amplitude $F_0/2$ corresponding to a maximum beam power of $P_0=\SI{20}{\micro\watt}$ is enough to excite motion near the phase transition. While exciting the motion with the focused laser beam, the ions were still illuminated with a spatially extended cooling beam. This beam damps the dynamics introduced via the amplitude modulation. Typical powers of the cooling beam while conducting spectroscopy are \SIrange{200}{300}{\micro\watt}.

\subsubsection{Numerical calculations}\label{sssec:NumericalCalculations}

\begin{figure}
	\centering
	\includegraphics[width=0.43\textwidth]{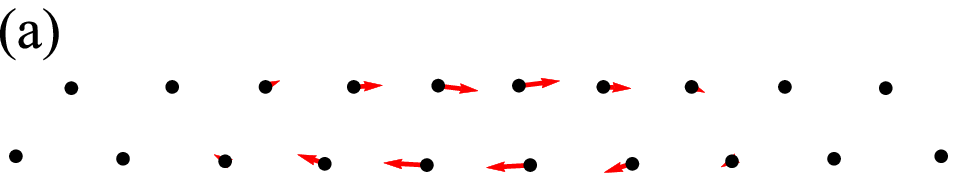}\\
	\includegraphics[width=0.45\textwidth]{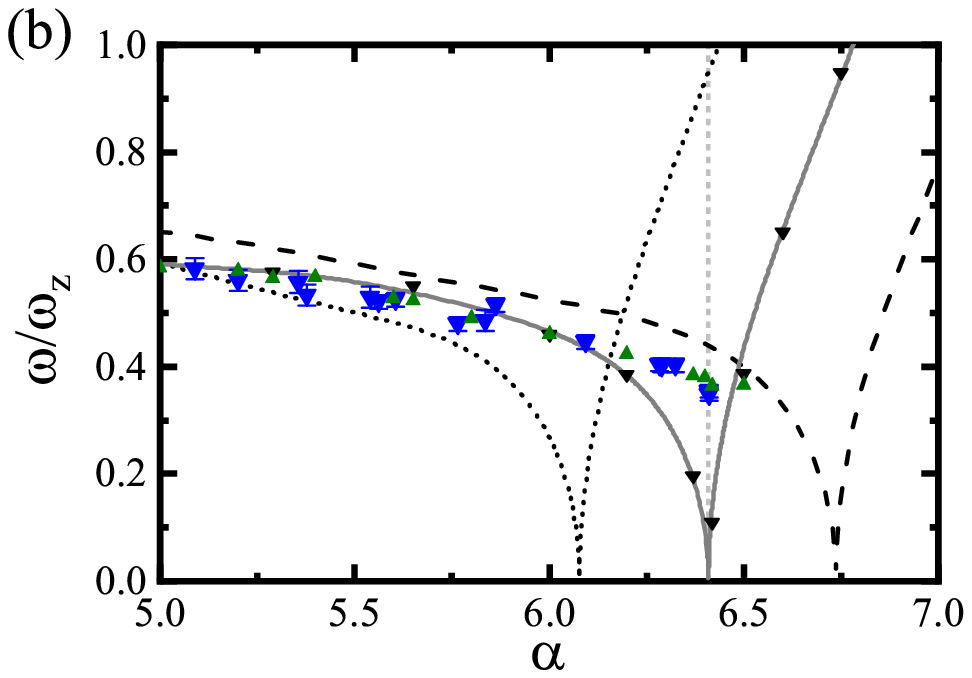}
	\caption{(a) Numerically obtained mode vector. Amplitudes are in arbitrary units and values smaller than \SI{10}{\percent} of the maximum amplitude are not shown. (b) Mode frequency against the control parameter $\alpha$. Experimental data taken in the high-precision trap for a 30 ion crystal is plotted as blue triangles. Error bars are given by the uncertainties of the measured common mode and defect mode frequency. Numerical calculations at $T=\SI{0}{\kelvin}$ are shown for $N=28$ (black dotted line), $N=30$ (gray solid line) and $N=32$ (black dashed line). Frequencies obtained via the Fourier transformation of molecular dynamics simulation are shown for $T=\SI{5}{\micro\kelvin}$ (black triangles) and $T=\SI{1}{\milli\kelvin}$ (green triangles). All angular frequencies are plotted in units of the axial common mode angular frequency $\omega_z=2\pi\times\SI{25.6(2)}{\kilo\hertz}$.}\label{fig:FrequencyNumerics}
\end{figure}

Firstly, we look at the expected mode frequency in dependence on the control parameter $\alpha$. For this, we analyze the total potential energy of $N$ harmonically trapped ions in two dimensions:
\begin{eqnarray}
V=&\sum_{i=1}^{N}\frac{m}{2}\left(\omega_x^2 x_i^2+\omega_z^2 z_i^2\right)\nonumber \\
&+\frac{e^2}{4\pi\varepsilon_0}\sum_{i=1}^{N}\sum_{j<i}\left|\mathbf{r}_i-\mathbf{r}_j\right|^{-1},\label{eq:totalPotential2D}
\end{eqnarray}
where $x_i$ and $z_i$ are the coordinates of $i$th ion in the coordinate systems defined in \Fref{fig:model_system}(a); $\omega_x$ and $\omega_z$ are the harmonic trapping frequencies; and $\bi{r}_i\equiv \{x_i, z_i \}$. The third dimension is neglected, due to the anisotropy of the radial secular frequencies. We obtain the frequency of the topological defect mode (and any other mode) by numerically diagonalizing the Hessian matrix
\begin{equation}
	H_{i,j}=\left.\frac{\partial^2 V}{\partial q_i q_j}\right|_{\mathbf{q}\left(0\right)} ,\label{eq:HessianMatrix}
\end{equation}
where $V$ is the total potential energy given by \Eref{eq:totalPotential2D}, $q_i$ are the degrees of freedom and $\mathbf{q}\left(0\right)$ is a given equilibrium configuration. The stable equilibrium configuration, $\mathbf{q}\left(0\right)$, is obtained by numerically solving the equations of motions under high damping. There are $2N$ vibrational normal modes in a two-dimensional ion crystal with $N$ ions. The $k$th eigenvector of $H_{ij}$ is the $k$th normal mode of the crystal $\bmu_k$, and the corresponding eigenvalue $\lambda_k$ determines the mode frequency via the relation $\omega_k=\sqrt{\lambda_k/m}$. The localized defect mode has the lowest frequency for $\alpha<7$, and it drives the symmetry breaking transition. \Fref{fig:FrequencyNumerics}(b) shows the shear mode frequency as a function of $\alpha$ for different numbers of ions $N$. The frequency tends to zero near the critical point $\alpha_c\approx 6.41\,(6.08, 6.74)$ for $N=30\,(28, 32)$, while increasing both above and below $\alpha_c$, because of the finite size of the system. The critical point $\alpha_c$ depends on the ion number $N$ due to the increase in charge density with more $N$. The more ions are present, the stronger is the repulsive Coulomb force, and therefore the more radial confinement, and subsequently higher $\alpha$, is necessary to bring the chains close enough together to observe the AT transition. In a previously installed prototype trap \cite{Pyka2014}, we observed the dependence of the defect mode frequency on $N$, but with consistently lower frequencies, than the calculations predict.
Near the AT transition we observe a difference between the measured and calculated frequencies in both the high-precision and the prototype trap.
This discrepancy stems mainly from the missing temperature effects in the numerical calculations.

\begin{figure}
	\centering
	\includegraphics[width=0.5\textwidth]{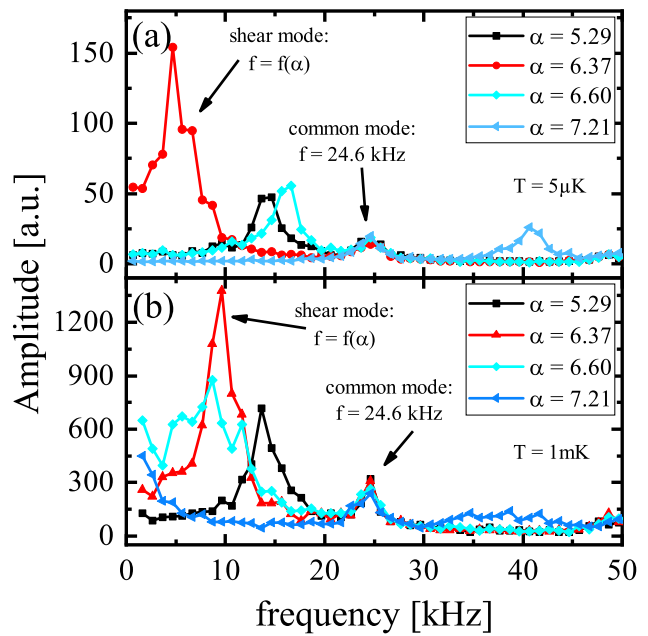}
	\caption{Spectra obtained via Fourier transformation of MD simulations. (a) Spectrum for $T=\SI{5}{\micro\kelvin}$ for different $\alpha$. Common mode, independent of $\alpha$, visible at \SI{24.6}{\kilo\hertz}. The frequency of the defect mode depends on $\alpha$. (b) Spectrum for $T=\SI{1}{\milli\kelvin}$. The frequency spread of the defect mode becomes broader the closer the system is to $\alpha_c$. Above the transition ($\alpha=6.6$ and $\alpha=7.21$) a range of several \si{\kilo\hertz} with comparable amplitude is observed. Therefore, no single harmonic oscillation has an observable amplitude. }\label{fig:FourierTransform}
\end{figure}
In order to investigate the effect of a finite crystal temperature, we conduct three-dimensional molecular dynamics simulations of an unperturbed crystal (i.e. without laser excitation, but with cooling). These simulations solve the Langevin equations, utilizing an impulse integrator algorithm \cite{Skeel2002}. The total potential energy of a three-dimensional $N$ ion system is
\begin{eqnarray}
V=&\sum_{i=1}^{N}\frac{m}{2}\left(\omega_x^2 x_i^2+\omega_y^2 y_i^2+\omega_z^2 z_i^2\right)\nonumber \\
&+\frac{e^2}{4\pi\varepsilon_0}\sum_{i=1}^{N}\sum_{j<i}\left|\mathbf{r}_i-\mathbf{r}_j\right|^{-1},\label{eq:totalPotential3D}
\end{eqnarray}
where $x_i$, $y_i$ and $z_i$ are the coordinates of $i$th ion $\omega_x$, $\omega_y$, $\omega_z$ are the harmonic trapping frequencies; and $\bi{r}_i\equiv \{x_i, y_i, z_i \}$. The equation of motion for the $j$th ion in one dimension is
\begin{equation}
m\ddot{\chi}_j+\nabla_{\chi} V+\eta\dot{\chi}_j=\xi_{\chi j}\left(t\right),\label{eq:Langevin}
\end{equation}
where $\xi_{\chi j}\left(t\right)$ is a stochastic force, $\chi$ is $x$, $y$ or $z$, $\eta$ is a damping term from laser cooling and $V$ is the total potential energy from \Eref{eq:totalPotential3D}. The stochastic force fulfills the following two conditions:
\begin{eqnarray}
 \left\langle\xi_{\chi j}\left(t\right)\right\rangle&=&0\label{eq:AverageEnergy}\\
 \left\langle\xi_{\chi j}\left(t\right)\xi_{\gamma k}\left(t'\right)\right\rangle&=&2\eta k_B T\delta_{\chi\gamma}\delta\left(t-t'\right) ,\label{eq:FluctuationDissipation}
\end{eqnarray}
where $\gamma$ is $x$, $y$ or $z$, $j$ and $k$ are ion indices and $\left\langle\dots\right\rangle$ indicates time averaging. \Eref{eq:FluctuationDissipation} is the fluctuation-dissipation relation that ensures that the system undergoing the stochastic dynamics equilibrates at temperature $T$ \cite{Pyka2013}. In these simulations we look at crystals with topological defects for different temperatures $T$, each time letting the system evolve freely for \SI{10}{\milli\second}. We then analyze the resulting axial trajectory of the 15th ion, $z_{15}\left(t\right)$, because this ion has one of the highest mode amplitudes for the localized defect mode in axial direction. Taking the Fourier transform of this trajectory 
\begin{equation}
\hat{z}_{15}\left(\nu\right)=\mathcal{F}\left(z_{15}\left(t\right)\right)\label{eq:FourierTransform}
\end{equation}
yields the spectrum of the harmonic motions excited in the crystal dynamics. \Fref{fig:FourierTransform}(a) shows the spectrum for a crystal at $T=\SI{5}{\micro\kelvin}$ for different trapping ratios $\alpha$. The amplitude peak at a frequency of \SI{24.6}{\kilo\hertz} is due to the excitation of the axial common mode, for which all ions oscillate in phase, and the other amplitude peaks are due to excitation of the topological defect mode at different corrugations. The dependence of the mode frequency on $\alpha$ at \SI{5}{\micro\kelvin} is shown with black triangles in \Fref{fig:FrequencyNumerics}(b). These results agree with the frequencies obtained from the diagonalization of the Hessian matrix. The observed Fourier spectrum changes, when $T$ is increased. At experimentally realistic temperatures of \SI{1}{\milli\kelvin} the Fourier spectrum close to, and above, the transition changes, as can be seen in \Fref{fig:FourierTransform}(b). The common mode can still be resolved independently of the trapping ratio, but only below $\alpha_c$ we observe distinct peaks for the defect mode in the spectrum. Additionally, these peaks are visible at different frequencies in comparison to the results for $T=\SI{5}{\micro\kelvin}$. For $\alpha>\alpha_c$ a broad range of frequencies with nearly identical amplitudes is seen (for $\alpha=6.6$ near \SI{10}{\kilo\hertz} and for $\alpha=7.21$ near \SI{40}{\kilo\hertz}), indicating that no single harmonic oscillation is present. This can be explained by the PN potential barriers, which are in the \si{\milli\kelvin} range, see \Fref{fig:SymBreak}(c), and small enough for the defect to overcome. During the movement between the minima, the non-linear parts of the PN potential are sampled and many frequency components are part of the dynamics, explaining why the mode of the defect could not be excited with an observable amplitude in the experiment for $\alpha>\alpha_c$. Furthermore, we find that the frequencies extracted from the Fourier analysis for $T=\SI{1}{\milli\kelvin}$ agree with the experimentally observed frequencies, see \Fref{fig:FrequencyNumerics}(b). This indicates that the frequency measurements were limited by the finite Doppler cooling temperature of the ${}^2$S$_{1/2}\rightarrow{}^2$P$_{1/2}$ transition.

\section{Symmetry breaking by external forces}\label{sec:Environment}
\begin{figure*}[btp]
	\centering
	\includegraphics[width=0.80\textwidth]{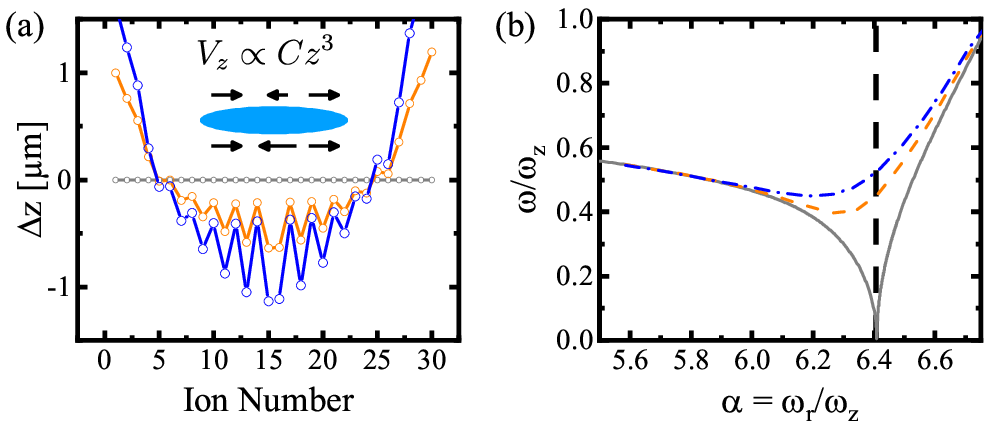}
	\includegraphics[width=0.80\textwidth]{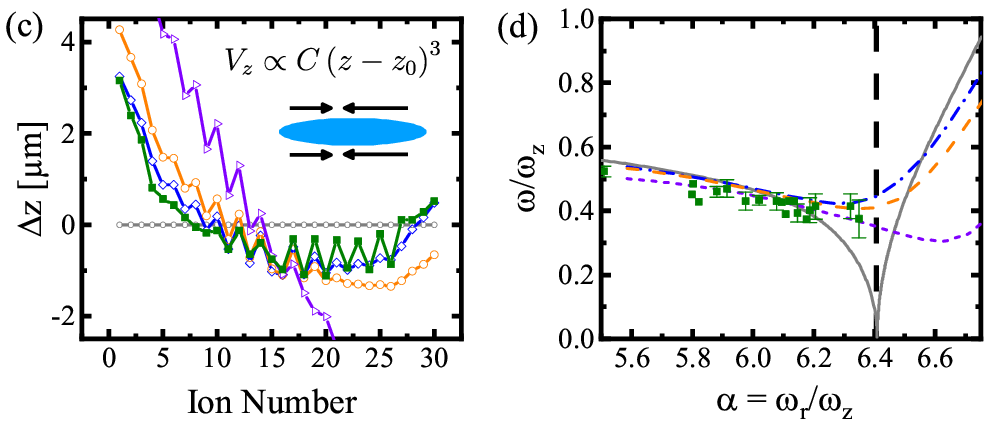}
	\caption{Influence of anharmonic potentials on the crystal symmetry and frequency dependence of the defect mode in comparison to experimentally observed crystal configurations and frequencies (green filled squares). The blue ellipsoids sketch the deformation of the axial ion positions over a crystal. (a) Difference in axial position $\Delta z$ between the modified equations with a third order term in the axial trapping potential and the unperturbed ponderomotive solution. Shown are numerical simulations for $C=0$ (i.e. no perturbation, gray empty dots) $C=-\SI{170}{\per\metre}$ (orange empty dots) and $C=-\SI{300}{\per\metre}$ (blue empty dots). $\alpha=5.51$. The ions on the left side are closer to the center, while the ions on right side are moved away by this perturbation. (b) Defect mode frequencies against control parameter for an anharmonic axial potential. Color scheme identical to (a). The modified crystal leads to a non vanishing of the mode frequency at $\alpha_c$. (c) Change in the axial coordinate for a spatial shift of the anharmonic potential with respect to the harmonic potential. Numerical simulations for $z_0$ of $0$ (i.e. no perturbation, gray empty dots), \SI{30}{\micro\metre} (blue empty diamonds), \SI{50}{\micro\metre} (orange empty circles) and \SI{150}{\micro\metre} (violet empty triangles), as well as experimental data (green squares) taken in the prototype trap. $\alpha=5.51$. All calculations have been carried out with $C=-\SI{300}{\per\metre}$, expect for $z_0=0$, where $C$ was also $0$. (d) Defect mode frequencies against control parameter for a spatial shift between harmonic and anharmonic part of the axial potential. Color scheme identical to (c). No vanishing of the mode frequency is observed and the position of the minimal frequency is shifted.}\label{fig:AnharmonicPotential}
\end{figure*}
In this section we investigate several external forces on ion Coulomb crystals and how they change the crystal structure and subsequently the mode frequencies. This is of interest for directed transport of kink solitons \cite{Brox2017} and the frictional properties, as a change in the soliton mode frequency implies different interaction energies between the chains. In a previously used prototype trap \cite{Pyka2014}, we observed asymmetric crystal configurations and measured lower frequencies of the localized defect mode, than the calculations predict. Therefore, we investigate here trap related forces, such as those stemming from higher order terms in the axial trapping potential and axial micromotion, as well as forces due to laser illumination of the ions. While all effects change the crystal and hence the position of the topological defect, the latter effect offers the possibility to move the defect at will through the crystal.

\subsection{Anharmonicities of the trapping potential}
As the manufacturing process of an ion trap is limited by its tolerances, the trap potential will contain non-quadratic contributions \cite{Gudjons1997}.

Here, we investigate the influence of a third order term in the axial trapping potential on the crystal structure and the frequency dependence on the corrugation. This is done numerically by changing the potential term for the $z$ direction:
	\begin{equation*}
		m\omega_{z}^2 z^2 \rightarrow m\omega_{z}^2\left(z^2 + C\cdot z^3\right),
	\end{equation*}
where $C$ determines the relative strength of the third order term. The addition of the anharmonic term to the potential perturbs the equilibrium positions of the ions. We define the difference of the axial positions of a modified crystal $z_{j,\text{mod}}$ and the axial positions of an unperturbed crystal $z_{j,\text{pond}}$ as $\Delta z_j = z_{j,\text{anharmonic}} - z_{j,\text{pond}}$, where $j$ is the ion index. We plot the difference in \Fref{fig:AnharmonicPotential}. $C$ was chosen such that the resulting equilibrium positions fit best to the positions observed in the prototype trap. The modified confining potential has a higher curvature on one side of the crystal, while the other side has a lower curvature in comparison to the ideal case. One side of the crystal is therefore pushed closer to the center, while the other side is shifted away from it. This change in the axial positions influences the interaction between the ions and thus modifies the dependence of the localized defect mode frequency on $\alpha$, see \Fref{fig:AnharmonicPotential}(b). In particular, in the presence of the third order term in the axial potential, the mode softening is less pronounced or not present at all, depending on the strength of $C$.

Up to now we have assumed, that the minimum of the second order term and the saddle point of the third order term are at the same axial position. As far as we know, there is no reason that this is necessary. To match the experimental crystal configuration to the simulation, we include a differential shift $z_0$ in the position of $z^2$ and $z^3$ terms in regard to the crystal structure and the mode frequency, with results shown in \Fref{fig:AnharmonicPotential}(c,d). The values of $z_0$ and $C$ are fitted to the observed crystal configurations, which we also plot in \Fref{fig:AnharmonicPotential}(c). Shifting the $z^3$ term by only \SI{30}{\micro\metre} relative to the $z^2$ term yields a significantly distorted crystal configuration, in which the left side is pushed towards the center by a few \si{\micro\metre}, while the right side is only slightly perturbed. In the frequency dependence we observe not only a less pronounced soft mode, but also a shift of the minimum of the localized mode frequency towards higher values of the control parameter $\alpha$. As it is possible to fit the simulations with anharmonic terms closely to observed crystals in the prototype trap, it is plausible, that the potential of the prototype trap included significant higher order terms. This is most likely due to the manufacturing process, as the prototype trap was milled, resulting in slightly asymmetric trap segments.

Both the anharmonic potential itself and the shift from the harmonic part lead to an intrinsic symmetry breaking and a change of the defect mode frequency near the transition. It is therefore necessary to suppress anharmonic potential terms in an ion trap, for experiments, that require high symmetry of the crystal. The higher order terms can be minimized by the design and careful construction of the trap.

\subsection{Micromotion}
\begin{figure}
	\centering
	\includegraphics[width=0.405\textwidth]{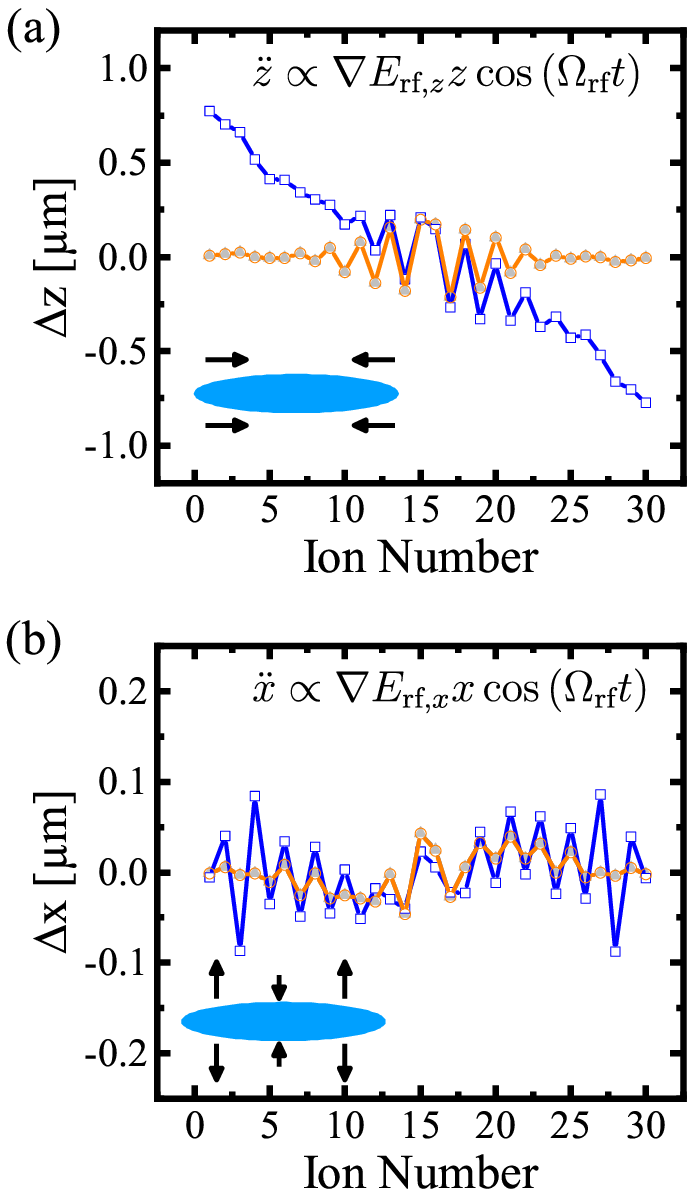}
	\caption{Influence of micromotion with and without axial rf gradients. (a) Axial difference of ion positions for crystals calculated from the Mathieu equations and the ponderomotive approximation. The numerical calculations for an axial rf field gradient $\nabla E_{\text{rf},z}$ of \SI{6.4d6}{\volt\metre^{-2}} (blue empty squares), \SI{6.4e4}{\volt\metre^{-2}} (orange empty circles, high precision trap) and for no axial micromotion (gray triangles) are shown.  The result for the high-precision trap is identical to the result for no axial micromotion. (b) Radial difference between the ion positions. Color code identical to (a). The blue ellipsoids sketch the deformations of the axial and radial ion positions over a crystal for $\nabla E_{\text{rf},z}= \SI{6.4d6}{\volt\metre^{-2}}$. For both graphs: $\alpha=5.51$.}\label{fig:micromotion}
\end{figure}
As all investigations take place in a 2D crystal, where most ions are situated in a non-zero rf field, the micromotion induced by the oscillating trapping field changes the structure and the dynamics of the ion crystal in comparison with the ponderomotive approximation \cite{Landa2012}. Furthermore, due to the finite size of the segmented linear Paul trap there are electric field gradients from the rf fields along the axial direction. This additional gradient induces axial micromotion, which might also influence the crystal.

The dynamics of ions trapped in a rf field are described by the Mathieu equations \cite{Ghosh1995}. Solving these equations for a 30 ion crystal containing a defect at a radial trapping frequency $\omega_x=\SI{132}{\kilo\hertz}$ and comparing it to the ponderomotive approximation, reveals differences $\Delta z_j = z_{j,\text{Mathieu}}-z_{j,\text{pond}}$ of less than \SI{200}{\nano\metre} in the $z$ coordinates of the ions. If axial micromotion with an electric field gradient $\left|\nabla E_{\text{rf},z}\right|$ of \SI{2}{\percent} of the radial electric field gradient $\left|\nabla E_{\text{rf},r}\right|=\SI{3.2e8}{\volt\metre^{-2}}$ is also present, then the crystal is axially compressed, see \Fref{fig:micromotion}(a). The further away an ion is from the center of the trap, the further its position is changed by the micromotion. Because the rf gradient changes sign in the center of the trapping region it acts like an additional axial potential, as long as the crystal is in the center of the trapping region. In the prototype trap the axial micromotion has an electric field gradient of ${\left|\nabla E_{\text{rf},z}\right| = \SI{1.9d6}{\volt\metre^{-2}}}$, which is roughly \SI{0.6}{\percent} of the radial electric field gradient $\left|\nabla E_{\text{rf},r}\right|$ \cite{Pyka2014} and in the precision trap we observe only an electric field gradient of ${\left|\nabla E_{\text{rf},z}\right|\approx \SI{6.4d4}{\volt\metre^{-2}} \approx 2\cdot 10^{-4}\cdot \left|\nabla E_{\text{rf},r}\right|}$ along the axial direction \cite{Keller2017}. For both traps, the calculated differences $\Delta z_j$ are nearly identical to the simulations without any axial micromotion, making the effect negligible. In the radial direction the effect of axial micromotion is small in comparison. In \Fref{fig:micromotion}(b) we plot the difference in the $x$ coordinate. For a high axial micromotion of $\SI{2}{\percent}\cdot \left|\nabla E_{\text{rf},r}\right|$ the difference in radial coordinates $\Delta x_j$ due to the axial micromotion are less than \SI{100}{\nano\metre}. Using the axial electric field gradients from both traps, we again do not observe a difference to the simulations without any axial micromotion.

The mode frequencies were not calculated for this effect, as the simulation using the full Mathieu equations does not yield a static solution needed for the Hessian matrix approach. It is possible to calculate the normal mode spectrum using the Floquet-Lyapunov approach \cite{Kaufmann2012,Landa2012}, but this is outside of the scope of this paper. The influence of micromotion is only relevant for traps with an axial rf field gradient higher than \SI{0.6}{\percent} of the radial rf field gradient. This gradient can only be suppressed by the design and careful manufacturing of the trap \cite{Herschbach2012}.

\subsection{Differential light forces}\label{ssec:lightforces}
Illuminating ions with near resonant light will lead to an average force acting on them. These light forces will in general modify the crystal structure, but as long as the illumination is uniform on the crystal, they will only shift the center of mass. An illumination that leads to an axial differential force between the two ion chains will move the topological. The position of the topological defect is sensitive to forces in the \SI{e-21}{\newton} level.

\begin{figure}
	\centering
	\includegraphics[width=0.405\textwidth]{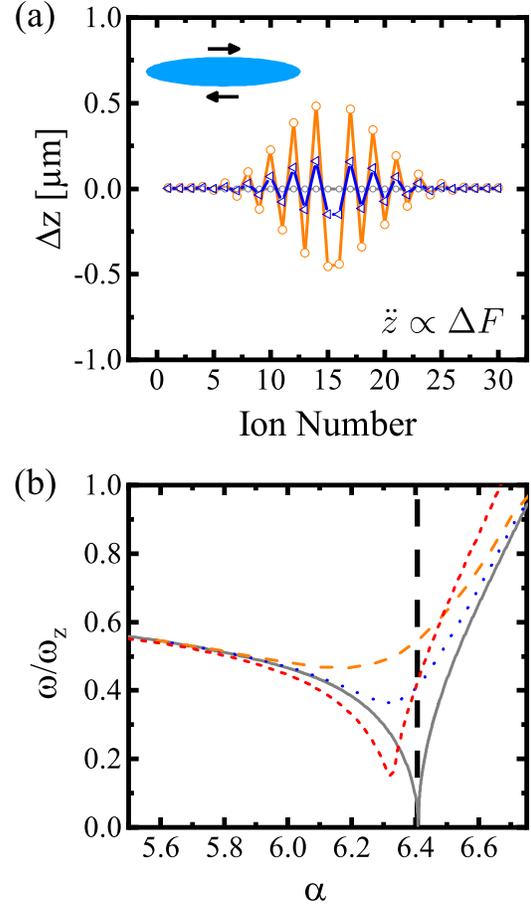}
	\caption{Influence of light forces. (a) Difference in axial position from the ideal crystal $\Delta z$ for differential light forces $\Delta F$ of \SI{0.9e-21}{\newton} (blue diamonds) and \SI{2.7e-21}{\newton} (orange circles) from numerical simulations. For comparison an ideal crystal is shown as gray circles. $\alpha=5.51$. The blue ellipsoid sketches the deformation of the axial ion positions over a crystal. (b) Frequency dependence on the interaction strength for different applied differential forces. Shown are results for $\Delta F=\SI{0.9e-21}{\newton}$ (dotted blue line), \SI{2.7e-21}{\newton} (orange dashed line) and \SI{9.8e-21}{\newton} (red short-dash line).}\label{fig:lightforces}
\end{figure}
In the experiments we use two focused cooling laser beams, one for cooling and detection and another for vibrational mode excitation, see \Sref{ssec:experiment} and \Sref{ssec:Spectroscopy}. A typical zigzag configuration in our setup has a radial separation of the ion chains of approximately \SI{15}{\micro\metre} and has a \SI{45}{\degree} angle to the beam wavefront, leading to an apparent separation of \SI{10.6}{\micro\metre}, which is not much smaller than the radial beam waist of approximately \SI{80}{\micro\metre}. Any alignment of the radial beam, that results in non-uniform illumination between the ion chains, will lead to a differential force {$\Delta F = F_1 - F_2 \ne 0$} along the axial direction, where $F_{1,2}$ is the force along $z$ acting on chain 1 and chain 2, respectively. From numerical simulations we find, that such a differential force will move the topological defect inside the crystal. This can be actively used, to investigate charge or information transport in ion Coulomb crystals \cite{Landa2010}, as well as further investigations into nanofriction of two sliding chains. In \Sref{sec:KinkMovement} we investigate in detail, what such a movement entails.

When applying differential forces to the crystal, the localized kink mode frequency near the phase transition will change. For different $\Delta F$ we plot the change in the axial crystal structure in \Fref{fig:lightforces}(a) and the expected frequency dependence on $\alpha$ in (b). This can be understood by examining the Peierls-Nabarro potential, shown in \Fref{fig:SymBreak}(c). The potential is overall confining due to the inhomogeneity of the crystal, and it contains PN barriers above the AT transition. The overall positive curvature of the PN potential explains why a finite frequency below the AT transition exists. This positive curvature is balanced by the negative curvature of the emerging PN barrier at $z=0$ when the system is near the phase transition, resulting in near zero frequency of the localized mode. However, when differential forces are applied, the kink soliton is shifted away from the point, where a PN barrier is formed. This leads to the observed non-zero harmonic frequency. Increasing $\Delta F$ further, results again in a soft mode behavior of the defect frequency, see \Fref{fig:lightforces}(b) for $\Delta F=\SI{9.8e-21}{\newton}$.

In our setup, working with a waist of \SI{80}{\micro\metre}, the maximum differential light force due to a laser beam on a zigzag crystal is achieved, if the radial crystal center is situated near the maximum slope, i.e. at $z \approx \pm w/2$, of the Gaussian intensity distribution $I\left(z\right) = I_0\exp\left(-2z^2w_0^{-2}\right)$. As an example, if a crystal is placed on the slope, $\Delta F$ around \SI{2.1e-21}{\newton} can be reached with an effective power $P_e = P_0/2$ of \SI{10}{\micro\watt}. From the numerical simulations we gather, that such forces result in a finite minimum of the mode frequency. For the frequency measurement we needed to have a slight differential light force in the laser beam, so that it can excite the axial shear mode. For this we adjusted the intensity maximum of the laser beam to one of the chains, resulting in roughly \SI{6}{\percent} intensity difference between the rows. We expect the resulting differential force to increase the minimal frequency to $0.32\omega_z$. This is currently below the measured frequency and is not yet limiting the experiments. Since a lower mode frequency needs smaller $\Delta F$ to excite the motion, this future limitation can be prevented by reducing laser power or reducing the intensity difference between the chains.

To avoid distortion of the crystal structure it is therefore necessary to symmetrically center the laser beam onto the ions, minimizing differential light forces.

\FloatBarrier
\section{Motion of the topological defect}\label{sec:KinkMovement}

In this section we discuss the motion of the topological defect, when the chains are pushed by differential forces introduced in the previous section. We will look at the motion for both the horizontal and vertical kink, and compare their behavior.

\begin{figure}[t]
	\centering
	\includegraphics[width=0.45\textwidth]{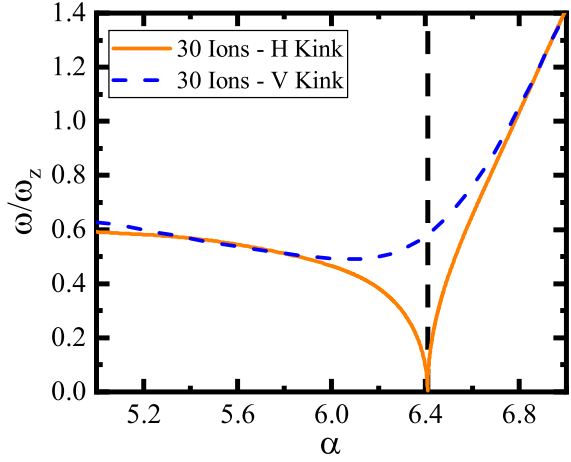}
	\caption{Defect mode frequency for different defect types. Plotted are the relation between mode frequency and control parameter $\alpha$ for the horizontal (orange solid line) and the vertical (blue dashed line) defect.}\label{fig:kinktypes}
\end{figure}

\begin{figure}
	\centering
	\includegraphics[width=0.45\textwidth]{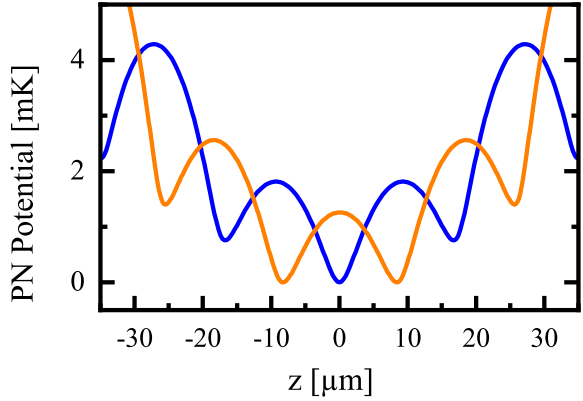}
	\caption{Peierls-Nabarro potential for both extended defects above $\alpha_c$. The PN potential for the horizontal defect (orange line, shows a PN barrier, at the center of the crystal. The PN potential for the vertical defect (blue line) shows no PN barrier at the center part of the crystal.}\label{fig:PNPotV}
\end{figure}

Starting with the frequency of the vibrational defect mode for the vertical kink we find, that in contrast to the horizontal topological defect, it does not exhibit a soft mode tending to zero, see \Fref{fig:kinktypes}, or a symmetry breaking at the critical point.  This can be explained by examining the PN potentials shown in \Fref{fig:PNPotV}. For $\alpha>\alpha_c$ the PN potential of the vertical kink has a minimum in the center of the trap, whereas the horizontal kink has a maximum. When the transition is crossed, the vertical kink does not move spontaneously, because the global minimum before the transition is still the global minimum after the transition.

\begin{figure}
	\centering
	\includegraphics[width=0.43\textwidth]{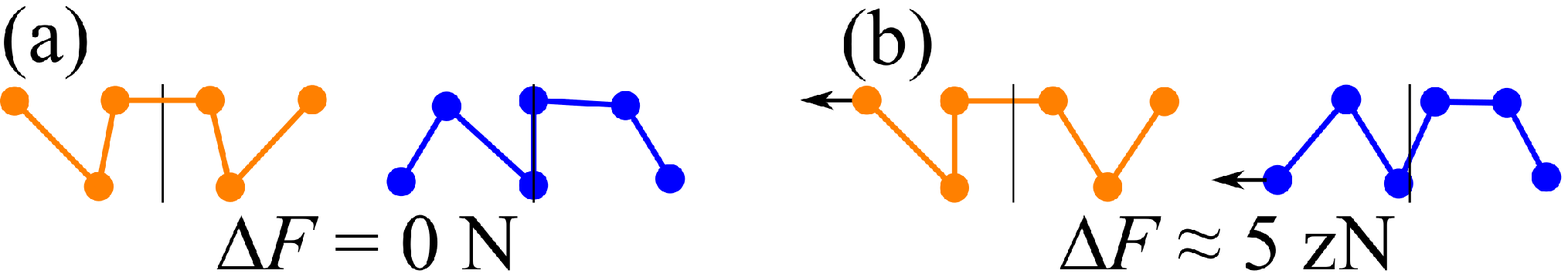}\\
	\includegraphics[width=0.45\textwidth]{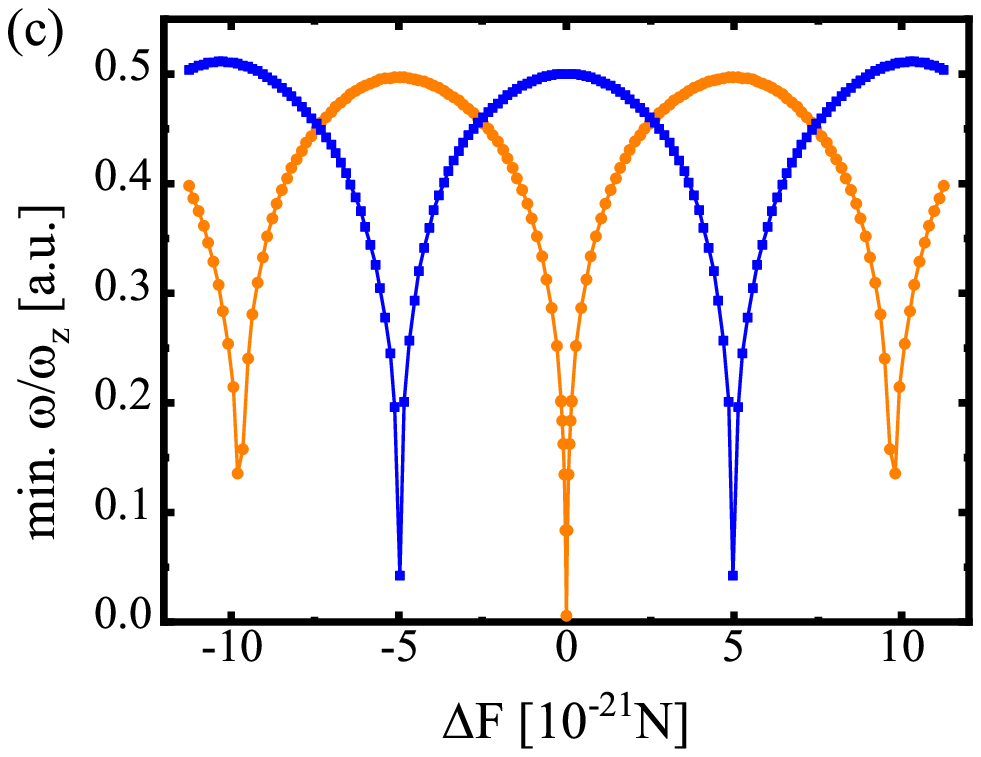}
	\caption{Shear forces move and change topological defects. (a) Local topological defect structure for a horizontal (left, orange) and a vertical defect (right, blue) for $\Delta F=0$. (b) Local defect structure for a horizontal and a vertical kink for $\Delta F\approx\SI{5e-21}{\newton}$. (c) Numerically obtained minimum of the soliton mode frequency in dependence of an applied force difference $\Delta F$ for the horizontal (orange circles) and vertical (blue squares) defect. For increasing $\Delta F$ the mode frequency of the horizontal defect becomes finite. A further increase leads to a reduction of the soft mode frequency again. The original crystal structure is reproduced at the next lattice site. This behavior repeats for each lattice site. The vertical kink shows the opposite behavior.}\label{fig:softmodeOscillation}
\end{figure}

By varying the force difference $\Delta F$ we push the kink through the crystal; the higher the $|\Delta F|$ the larger is the displacement of the kink from the center. We find that the local structure of a stable kink configuration changes periodically as a function of $\Delta F$, as is shown in  \Fref{fig:softmodeOscillation}(a) and (b). If $\Delta F$ is around \SI{5e-21}{\newton}, the local structure of a vertical defect looks like a horizontal defect. If the force is increased further to roughly \SI{10e-21}{\newton}, the initial local structure is reproduced, but one lattice period away from the initial position. Shown in \Fref{fig:softmodeOscillation}(c) is the soft mode frequency, i.e. the lowest frequency of the localized defect mode, against an applied $\Delta F$. In this relation a similar periodic behavior is observed.

We find, that the minimum mode frequency is low (tending towards zero), whenever the local structure looks like a H defect. The minimum is high (around $0.5 \omega_z$), when the defect is locally similar to the V defect. By starting with an H defect we observe the opposite behavior. This explains, why a differential force $\Delta F$ 'destroys' the soft mode. The local structure changes from an H defect to a V defect, which does not exhibit a soft mode.

Even though these calculations are done for static configurations, in a dynamic process, the ions will move along the same trajectory. Therefore, we conclude, that the motion of the extended kink in an ion Coulomb crystal is accompanied by a change in the nature of the defect, which can be seen both in its local structure and in the frequency of the vibrational defect mode. Note that the vertical and horizontal defect are still distinguishable by their number of ions per chain. Only locally they shift into each other.

\section{Discussion and Conclusion}\label{sec:Discussion}
In this work, we investigated atomic friction in self-organized ion Coulomb crystals with topological defects. We observed the Aubry-type transition from the sliding to the pinning regime and demonstrated experimentally and numerically the two main features of this transition, namely the symmetry breaking and the existence of a soft mode. The excitation of the low energy gapped mode of the topological kink soliton was experimentally demonstrated and its frequency was measured as a function of the distance between the two chains, which directly relates to the interaction energy between them. These findings are of direct relevance to research that aims to exploit topological defects for quantum information processing \cite{Landa2010}. The experimental observation of the mode softening at the Aubry-type transitions was limited by the finite temperature of the crystal, which leads to shifts of the mode frequencies. Molecular dynamics simulations demonstrate that lower temperatures, of the order of \si{\micro\kelvin}, are necessary to observe a strong softening of the defect mode.
The spectroscopic method, introduced in this work, for probing the vibrational mode of the topological defect with the help of amplitude modulated laser light is an alternative to typical excitation methods, such as parametric excitation via rf modulation \cite{Ibaraki2011,Brox2017}. It offers the advantage of being mode sensitive as the laser intensity distribution can be matched to fit a specific motional mode vector, providing a tool for investigating mode mixing in ion Coulomb crystals. 

We also numerically investigated the influence of external forces, present in ion trapping experiments, on the crystal structure and the kink soliton frequency. The sensitivity of the kink to these effects, can be used to characterize even small disturbances in ion trap experiments.
We were able to fit the experimental results, obtained in a previously used milled prototype trap \cite{Pyka2014}, closely to simulations which include an additional third order term in the axial potential. The additional non-linear term is most likely due to the fabrication method. In the new high-precision, laser-cut ion trap \cite{Keller2017}, an influence of higher order terms was not observed. We also investigated the influence of axial micromotion on the crystal configuration and found that, for experimentally observed values of micromotion, the ion positions are not changed in comparison to simulations using no axial micromotion. It is known that mode frequencies obtained from the ponderomotive approximation deviate by a few percent to frequencies obtained from the full dynamics \cite{Kaufmann2012}. However, as the relative experimental resolution of our method is currently around \SI{4}{\percent}, it is unlikely that we could resolve this frequency difference. In future experiments, the resolution can be improved, in which case  using the Floquet-Lyapunov approach to calculate the mode frequencies might be necessary.

With the help of differential light forces, of only a few \SI{e-21}{\newton}, the topological defect can be moved through the crystal. This effect is similar to force gradients due to anharmonic potentials, when directed transport was observed \cite{Brox2017}. We observed a periodic change in the local structure of the kink as a function of the applied external force, demonstrating 
that as the kink moves through the chain it alternates between two types of stable kinks, one with a gapped vibrational mode (vertical kink) and the other with a soft vibrational mode (horizontal kink).

In future experiments, the crystal temperature can be lowered to the \si{\micro\kelvin} regime, using dark resonances \cite{Morigi2000,Lechner2016,Scharnhorst2017}, Sisyphus cooling \cite{Ejtemaee2017} via polarization gradients or narrow transitions \cite{Pyka2014,Keller2017} in other ion species implanted into the crystal. Lower temperatures might enable friction experiments in the quantum regime \cite{Zanca2018}. Furthermore, investigations of crystals of higher dimensions or homogeneously spaced crystals can help in translating the results from our model system to solid state sliding systems. Here the versatility of ion traps can help, as either designed anharmonic potentials or ring traps \cite{Li2017} could be used to create such scenarios.


\FloatBarrier
\appendix
\ack We thank Jonas Keller for many useful discussions and valuable help with the experiment control. We thank Lars Timm for carrying out numerical calculations and for comments on the manuscript. This work was supported by the DFG through grant ME 3648/1-1.
\section*{References}
\bibliographystyle{iopart-num}
\bibliography{mylib}
 
\section*{Appendix}

\subsection*{Offset of the primary gap in the hull function}
In \Fref{fig:hullfunctions}(c) for $\alpha=6.6$, the central gap is offset from $z_{j,0}=0$ and $z_j=0$. This is due to broken mirror symmetry of the system. The topological defect is already in one of the central PN potential wells, see \Fref{fig:SymBreak}(c). The defect is not situated at the position of the PN barrier, and neither are the ions, for which we calculate the hull function. Before a slip can occur, the ions first need to move in either axial direction, which is the reason why the primary gap is not at $z_{j,0}=0$. Due to the inhomogeneity the localized defect moves towards one central minimum (towards the right in the figures), instead of the other, were the PN potential becomes steeper. This explains why the gap is not symmetric around $z_j=0$.
\subsection*{Simulation Overview}
Throughout this paper we used different numerical calculations depending on the issue at hand. The following list and \Tref{tab:SimulationOverview} gives a short description of each method and where it was used.

\begin{itemize}
	\item two-dimensional calculation using time-averaged rf potentials without temperature. This code solves the equations of motion of all ions in the ponderomotive approximation of the rf field under high damping to get the equilibrium positions $\mathbf{q}\left(0\right)$. $T=\SI{0}{\kelvin}$. It also allows for the diagonalization of Hessian matrix for a given $\mathbf{q}\left(0\right)$,  resulting in the mode spectrum of the crystal, and the calculation of the hull functions.
	\item three-dimensional calculation using time-averaged rf potentials with temperature. This code solves the Langevin equations for all ions in the ponderomotive approximation for the rf field in three dimensions under experimentally realistic damping. The temperature is introduced with a stochastic force. Vibrational mode frequencies are obtained with Fourier transformation of the resulting ion trajectories.
	\item three-dimensional calculation of time-dependent rf potentials without temperature. This code solves the Mathieu equations \cite{Ghosh1995} for all ions in three dimensions under damping. The average positions of the ions are taken as the equilibrium positions for comparison with the ponderomotive approximation.
\end{itemize}
\begin{table*}[!hbtp]
	\centering
	\begin{tabular}{|c|c|c|c|}
		\hline 
		Dimensions & EOM solved for & Temperature & Used in Figures \\ 
		\hline 
		2 & time-averaged rf potential & $\SI{0}{\kelvin}$ & \ref{fig:SymBreak}(b), \ref{fig:hullfunctions}(b), \ref{fig:FrequencyNumerics}(b), \ref{fig:AnharmonicPotential}, \ref{fig:lightforces}, \ref{fig:kinktypes}, \ref{fig:softmodeOscillation} \\ 
		\hline 
		3 & time-averaged rf potential & $>\SI{0}{\kelvin}$ & \ref{fig:FrequencyNumerics}(b), \ref{fig:FourierTransform} \\ 
		\hline 
		3 & time-dependent rf potential & $\SI{0}{\kelvin}$& \ref{fig:micromotion} \\ 
		\hline 
	\end{tabular}
	\caption{Overview showing in which Figure each type of numerical calculation is used. EOM: equations of motion}\label{tab:SimulationOverview}
\end{table*}

\end{document}